\documentclass[conference]{IEEEtran}
\usepackage{lineno,hyperref}
\usepackage{amsmath}
\usepackage{verbatimbox}
\usepackage{graphicx}
\usepackage{tabularx}
\usepackage[usenames, dvipsnames]{color}
\usepackage{rotating}
\usepackage{caption}
\usepackage{algorithm} 
\usepackage{algpseudocode} 
\usepackage{algorithmicx}

\usepackage{comment}

\usepackage{picins}
\usepackage{amssymb}
\usepackage{subfig}
\usepackage{amsmath}

\usepackage{algorithm}
\usepackage{algpseudocode}
\usepackage{amsthm}

\theoremstyle{plain}
\newtheorem{theorem}{Theorem}

\newtheorem{lemma}{Lemma}

\let\labelindent\relax
\usepackage{enumitem}
\setlist[itemize]{align=parleft,left=0pt..1em}

\usepackage{lipsum}
 
\usepackage{cite}

\def\bearn{\begin{eqnarray*}}
\def\eearn{\end{eqnarray*}}
\def\bear{\begin{eqnarray}}
\def\eear{\end{eqnarray}}
\def\barr{\begin{array}}
\def\earr{\end{array}}
\def\bmat{\left(\begin{array}}
\def\emat{\end{array}\right)}

\def\bd{\begin{definition}}
\def\ed{\end{definition}}
\def\bt{\begin{theorem}}
\def\et{\end{theorem}}
\def\be{\begin{center}\begin{equation}}
\def\ee{\end{equation}\end{center}}
\def\bc{\begin{corollary}}
\def\ec{\end{corollary}}
\def\bl{\begin{lemma}}
\def\el{\end{lemma}}
\def\br{\begin{remark}}
\def\er{\end{remark}}
\newtheorem{problem}{Problem}

\begin{document}

%




\title{Mobile Networks on the Move: Optimizing Moving Base Stations Dynamics in Urban Scenarios\vspace{-1.2em}}





\author{
    \IEEEauthorblockN{
        Laura Finarelli\IEEEauthorrefmark{1}\IEEEauthorrefmark{2}, 
        Falko Dressler\IEEEauthorrefmark{1}, 
        Marco Ajmone Marsan\IEEEauthorrefmark{3}, 
        Gianluca Rizzo\IEEEauthorrefmark{2}
    }
    \IEEEauthorblockA{
        \IEEEauthorrefmark{1}TU Berlin, Germany, \{surname\}@ccs-labs.org; 
        \IEEEauthorrefmark{2}HES-SO Valais, Switzerland, \{name.surname\}@hevs.ch; \\
        \IEEEauthorrefmark{3}Institute IMDEA Networks, Spain, marco.ajmone@imdea.org
    }
}

\maketitle

\begin{abstract} 
Base station densification is one of the key approaches for delivering high capacity in radio access networks. However, current static deployments are often impractical and financially unsustainable, as they increase both capital and operational expenditures of the network. An alternative paradigm is the moving base stations (MBSs) approach, by which part of base stations are installed on vehicles. However, to the best of our knowledge, it is still unclear if and up to which point MBSs allow decreasing the number of static base stations (BSs) deployed in urban settings.
In this work, we start tackling this issue by proposing a modeling approach for a first-order evaluation of potential infrastructure savings enabled by the MBSs paradigm. Starting from a set of stochastic geometry results, and a traffic demand profile over time, we formulate an optimization problem for the derivation of the optimal combination of moving and static BSs which minimizes the overall amount of BSs deployed, while guaranteeing a target mean QoS for users. Initial results on a two-district scenario with measurement-based network traffic profiles suggest that substantial infrastructure savings are achievable. We show that these results are robust against different values of user density.
\end{abstract}

\vspace{-0.2in}
\section{Introduction}

Cellular networks will face significant technical and economic challenges soon, primarily driven by the escalating number of users and the growing demand for data traffic, especially in urban settings. According to the forecasting analysis in \cite{forecast}, the per-area data volume in future systems will increase by 1000 times, while the number of connected devices and the user data rate will increase by 10 to 100 times. To address this challenge, network densification, which was already exploited in 5G \cite{densification}, was proposed. However, this approach leads to an increase in both capital expenditure (CAPEX) and operational expenditure (OPEX) of the network. Furthermore, densification is strictly related to over-provisioning, i.e. worst-case dimensioning and deployment of network resources, to cope with traffic peaks and with the increasing variability of traffic over time. To reduce network costs and installations, other solutions have been explored as the deployment of small cells or Distributed Antenna Systems (DAS). Among all of them, a promising alternative lies in the dynamic mobile network paradigm \cite{dymonetproposal}. The moving small cell base stations (BSs) provide additional capacity, when and where needed, to the end users of an otherwise traditional RAN. Studies about the optimal positioning and other challenges related to the mobility of MBSs have already been investigated and solved as in \cite{optimalpos}. Existing studies confirm the adaptability of MBSs, which have, in contrast with UAVs, the potential to align with both spatial and temporal variations in the number of mobile users and the level of traffic they generate in urban scenarios \cite{AJMONEMobilenet}. However, so far it is unclear to which point the introduction of MBSs may help decrease the total amount of BS deployed in an urban scenario.\\ 
In this paper, we provide a novel framework for a first-order quantitative evaluation of these potential infrastructure savings. 
Specifically, we propose an analytical approach for determining the optimal number of BSs required to implement a hybrid network, i.e., that encompasses both fixed and mobile BS, which minimizes the total number of installed BSs while guaranteeing the desired Quality of Service (QoS) for all users. 
Our main contributions are:
\vspace{-0.05in}
\begin{itemize}
    \item We formulate a linear optimization problem to determine the optimal number of moving and static BS required in a given scenario to serve a given traffic demand with a target QoS;
    \item We propose an approach to compute CAPEX savings, which is based on the adopted patterns of traffic demand in each district of an urban scenario for a given time interval. By solving the optimization problem, it computes the overall amount of static infrastructure required in each district (with and without MBSs), as well as the overall minimum amount of MBSs which allows for satisfying the traffic demand in the whole city for the given time interval; 
    \item We apply our approach on a two-district scenario, for a first-order evaluation of the savings achievable with the MBSs paradigm. Results suggest that MBSs enable substantial savings (reaching a maximum of $21\%$ in the considered settings) in the total amount of deployed BSs. We show that these results are robust with respect to different values of mean user density in each district.
\end{itemize}
\vspace{-0.15in}
\section{System model and Assumptions}
\label{sec:System model}
We consider a finite area of the plane, on which BS and users are distributed according to homogeneous planar Poisson Point Processes (PPP). User devices consist of broadband (BB) terminals, though our approach can be extended with heterogeneous devices, such as UAVs or IoT equipment. We study our system over a finite time interval, which we denote as \textit{observation window}. This choice accounts for the fact that in realistic scenarios patterns of traffic demand and user mobility are typically periodic over different timescales (day, week, month...). Thus our system can be dimensioned and characterized over an observation window equal to a multiple of one of such periods. 
We assume the observation window to be partitioned in a set of equal-sized intervals, and let $j\in 1,..., J$ be the label of the $j-$th interval. The choice of the duration of each slot is based on a tradeoff between the accuracy of the analysis, and its computational complexity, among others.
As a performance metric, we use the per-bit delay, defined as the inverse of the short-term user throughput \cite{perbitdelay2017mahdian}. 
We assume users require a mean per bit delay $\tau_0$ 
and that all base stations transmit at the same power, which does not vary over time. We assume all BSs use the same wireless communication technology (e.g. mmwave) for the backhaul link which connects them to a static access point. Indeed, wireless backhauling is widely adopted in urban environments, particularly in small cell deployments, to avoid the costs and practical hurdles of cabled connections. Thus, for the sake of simplicity of analysis, in this preliminary work, we do not model the impact of backhaul link performance on user-perceived performance. 
We assume part of the BSs to be mobile, and we denote them as MBSs. In particular, we consider the ideal case in which MBS can move freely throughout the entire scenario to follow traffic demand, and we defer to a later work the inclusion of constraints to MBS movement relative to actual traffic flows among regions over the considered observation window.
We assume that the area being modeled represents an urban scenario partitioned into $Z$ regions. Each region models a distinct land use, with its pattern of traffic demand during the observation window.
With $\lambda^{z}_{b}$ we denote the intensity of the PPP of BSs in the region $z \in {1,..., Z}$. Similarly, in every interval $j$ and region $z$, with $\lambda^{z}_{u,j}$ and $\lambda^z_{m,j}$ we denote the intensity of the PPP of UEs and MBSs respectively.
 We impose that our system satisfies the \textit{closed system} assumption. That is, the mean total number of moving BSs over time in the whole area is constant and equal to $M$. That is, $\forall j$, $\sum_{z=1}^Z\lambda^{z}_{m,j}A_z=M$, 
where $A_z$ is the area of region $z$. 
Our channel model only takes into account distance-dependent path loss. We assume that \textit{random frequency reuse} is in place, with reuse factor $k$. We assume that UEs are associated with the BS that provides the largest SINR at the user location. We consider urban scenarios where the assumption of high attenuation typically holds. In these settings, as no fading is considered and all BSs use the same transmit power, assuming that users associate to the closest BS is a reasonable approximation \cite{Baccelli_stochasticgeometry}. We focus on the downlink of the wireless channel, though similar considerations can be made for the uplink channel. 
We assume each base station shares its time equally among all users associated with it, e.g. in a round-robin fashion. At each round, we assume the base station is idle for a fraction of time, whose duration is tuned to have the mean per bit delay perceived by each user coincide with the target value. Then the BS utilization is the fraction of time the BS is actively serving users.
To compute the performance of the system in terms of average per bit delay $\bar{\tau}_j^z$ perceived by a user in the region $z$ at time $j$, we adopt results from \cite{rizzo_energy_2018,swipt} that correlate the user's average ideal per bit delay (i.e. the per bit delay perceived by users when the serving BS utilization is equal to 1) with the density of active BSs $\lambda_b^z$ and of other connected users $\lambda_{u,j}^z$ in the area:
\begin{equation}\label{eq:tau}
\bar{\tau}_j^z=\int_0^{\infty} \biggl(\int_0^{\infty} \int_0^{2 \pi}e^{- \lambda_b^z A(r,x,\theta)}\lambda_{u,j}^z x d \theta d x \biggr) \frac{e^{- \lambda_b^z \pi r^2} \lambda_b^z 2 \pi r}{C(r,P,G,\bar{I})} dr
\end{equation}
with $A(r, x, \theta)$ given by  
$\pi x^2 - \bigg[r^2 \cos^{-1} 
\left(\frac{r +  x \sin(\theta)}{d(r, x, \theta)}\right)+x^2 \cos^{-1} \left(\frac{x + r \sin(\theta)}{d(r, x, \theta)}\right)- \frac{1}{2} (-(d(r, x, \theta)-x)^2+ r^2)^\frac{1}{2}((d(r, x, \theta) +x)^2-r^2)^\frac{1}{2} \bigg]
$ 
and $d(r,x,\theta) = \sqrt{x^2+r^2+2xr\sin(\theta)}$.
The capacity $C$ is modeled using Shannon's capacity law, i.e. $C(r,P,G,I) = (B/k) \log_{2}\left(1 + \frac{P  r^{-\alpha}}{N_0+I(r,k)}\right)$, where $\alpha$ is the attenuation coefficient and $N_0$ is the power spectral density of the additive white Gaussian noise and, the average interference is given by $ \bar{I}(r,k) = \frac{P 2\pi r^{2-\alpha}\lambda_b^z}{k(\alpha-2)}\frac{\bar{\tau}}{\tau^0}$
where $\frac{\bar{\tau}}{\tau^0}$ is the mean BS utilization \cite{rizzo_energy_2018}.

\vspace{-0.15in}
\section{Estimation of CAPEX savings}
\label{sec:problem_formualtion}
The first step of our approach for the estimation of the mean savings related to the integration of MBSs in the cellular network, 
consists of solving an optimization problem that determines, for each time slot and region, the minimum total density of base stations required to serve the given population of users with the target mean per bit delay.
\begin{problem}[\textit{Minimum 
total density of BS with target QoS}]\label{prob:0}
$$
\underset{\Lambda_{b}}{\text{minimize}}\
  \sum_{z=1}^Z \sum_{j=1}^J \lambda_{b,j}^z$$
\quad s.t., $\forall j \in \{1,...,J\}, z \in \{1,...,Z\}, \quad
\bar{\tau}_j^z \leq \tau_0, \quad U_j^z \leq 1$
\end{problem}

where $\Lambda_{b}(j,z)= \lambda_{b,j}^z$, $\bar{\tau}_j^z$ is given by equation \ref{eq:tau} and $U_j^z = \frac{\bar{\tau}_j^z}{\tau_0}$ is the average BS utilization in slot $j$ and region $z$. In a scenario with only static BSs, Problem 1 allows deriving the minimum density of static BSs in each region $z$ which allows satisfying the constraint on user-perceived QoS across the whole observation window. Problem 1 is nonlinear and non-convex, and it can be solved by a heuristic such as Genetic Algorithm. 
Problem 1 solutions are then inputs to a second optimization problem, for the derivation of the number of static and moving BS, in each time slot $j$ and region $z$, which allows serving a given user density $\lambda_{u,j}^z$  with the target mean per bit delay $\tau_0$, while minimizing the total amount of deployed BSs.
\begin{problem}[\textit{Optimal density of moving and static BS}]\label{prob:1}
\begin{equation}\label{eq:objectivefunction}
\underset{\Lambda_{m},\mathbf{\lambda}_{b}}{\text{minimize}}\     
    M + \sum_{z=1}^Z\lambda^{z}_{b} A_z 
\end{equation}
\quad Subject to, $\forall j \in \{1,...,J\}, z \in \{1,...,Z\}$,
\begin{align}
& \sum_{z=1}^Z\lambda^{z}_{m,j}A_z=M\\
&   0 \leq \lambda_{m,j}^z \leq \max_{j}(\lambda_{b,j}^z),\quad 0 \leq  \lambda_b^z\leq \max_{j}(\lambda_{b,j}^z) \label{positivity}\\
&    \lambda_{b,j}^z \leq \lambda_{b}^z + \lambda_{m,j}^z \label{minbs}
\end{align}
\end{problem}

 where $\mathbf{\lambda_b} = (\lambda_b^1,...,\lambda_b^Z)$ and the entries of matrix $\Lambda_{m}$ are given by $\Lambda_{m}(j,z) = \lambda_{m,j}^z$. Constraints \ref{positivity} ensure that there will not be negative values for MBS density and no more MBS than the maximum required to satisfy users over all the time slots. 
 Inequality \ref{minbs} guarantees to have, at least, for each time slot and region, a total density of BS larger than the minimum required to serve all users with the desired QoS, as derived in Problem 1. As Problem 2 is linear, it can be solved efficiently using the interior-point method.
The CAPEX saving estimation is computed as the difference between the total amount of base stations deployed in the static-only case, given by the sum over every region $z$ of $\max_{j}(\lambda_{b,j}^z)$, and the value of the cost function of Problem 2 at the optimum. 

\vspace{-0.2in}
\section{Numerical Results}
\label{numerical results}
\begin{figure*}[ht!]
     \centering
     \subfloat[\centering Traffic profiles]{\includegraphics[width=.3\linewidth]{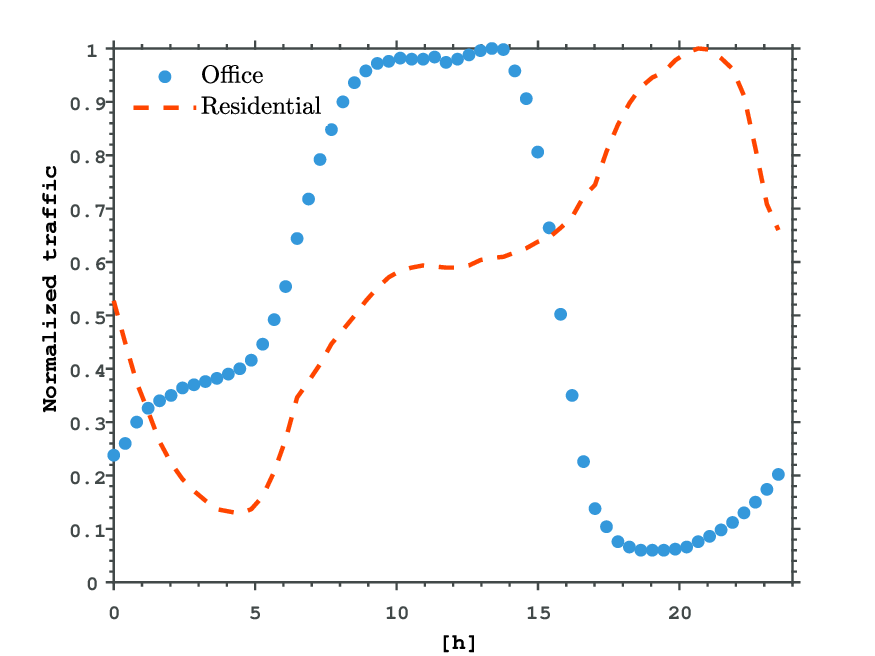}}
      \subfloat[\centering Office district]{ \includegraphics[width=.3\linewidth]{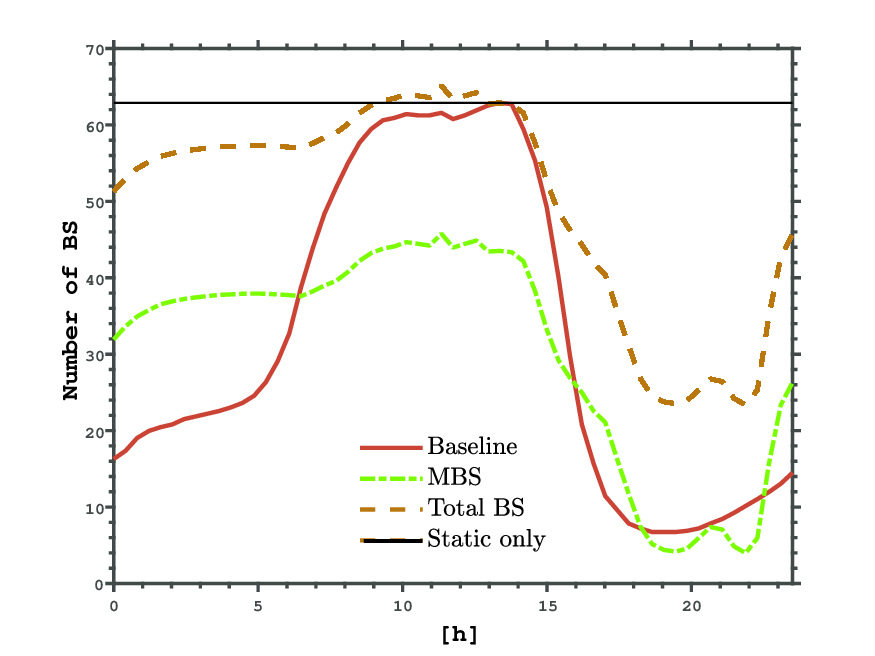}}
        \subfloat[\centering Residential district]{\includegraphics[width=.3\linewidth]{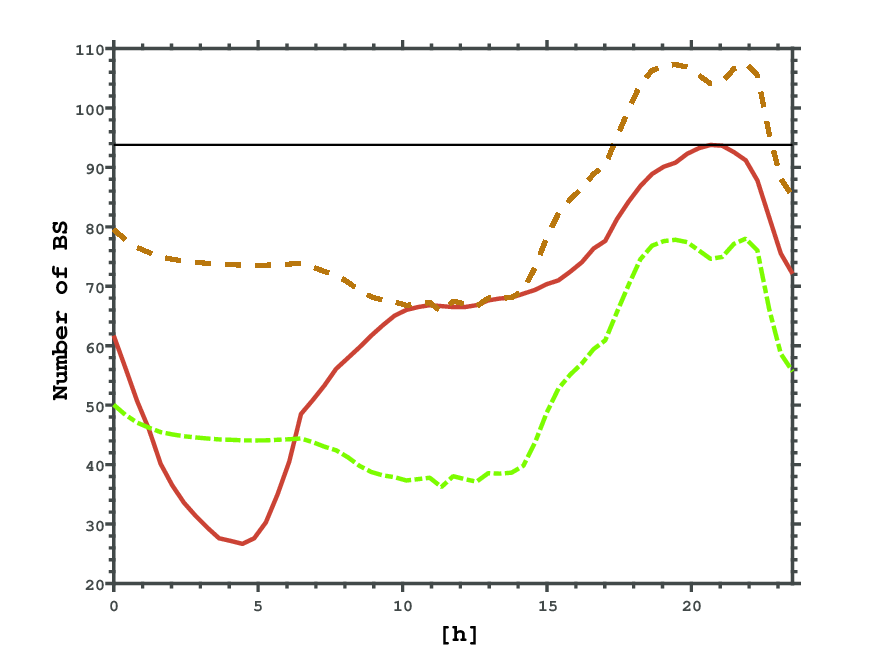}}
     \caption{(a) Normalized daily mobile traffic profiles in the resident (red) and office (blue) district; (b), (c): Optimal number of MBS (green), total number of BS (orange) with respect to the baseline (red), and installed BS in the static scenario (black) for the two regions over 24 hours. The total number of active base stations over the all area is given by the superposition of the orange dashed lines in the 2 plots.}  
     \label{comparisons}
     \vspace{-25pt}
\end{figure*}

To validate our approach, we considered an ideal urban scenario, composed of two regions with different patterns of network traffic demand, one modeling a residential area, and the other a business and office district. The daily network traffic profiles of the two areas are derived from operator data~\cite{9448695} (Fig. \ref{comparisons}a). Note that the peak in traffic of a region happens roughly at the same time at which there is a low in traffic in the other region. This suggests that a substantial amount of MBSs could be reused by moving them from one region to the other, to cater for peaks in traffic demand.  
 We partition the $24h$ observation window into 
 $60$ equal-sized intervals.
We consider a target per-bit delay of $10^{-5} \frac{s}{bit}$ (i.e. $100$ kbps, such as the data rate of a Skype call). 
Base stations work at a frequency of $1$ GHz with a bandwidth of $10$ MHz.
To estimate the mean number of active network users in every region, the network traffic profiles have been normalized and rescaled in such a way as to have the peak of active user density coincide with values that are considered reasonable in future dense urban scenarios \cite{worldpopulationreviewMilanPopulation}. Specifically, we set it to $10000$ and $1000$ users$/km^2$ for the business and residential districts respectively. 
Unless otherwise stated, we assumed a ratio between the areas of the office and residential regions (denoted as \textit{area ratio}) of $10$. We have considered scenarios in which the residential area is always larger than the office area of $1km^2$, as is common in 
many urban settings.\\
\begin{figure*}[ht!]
     \centering
        \subfloat[]{\includegraphics[width=.3\linewidth]{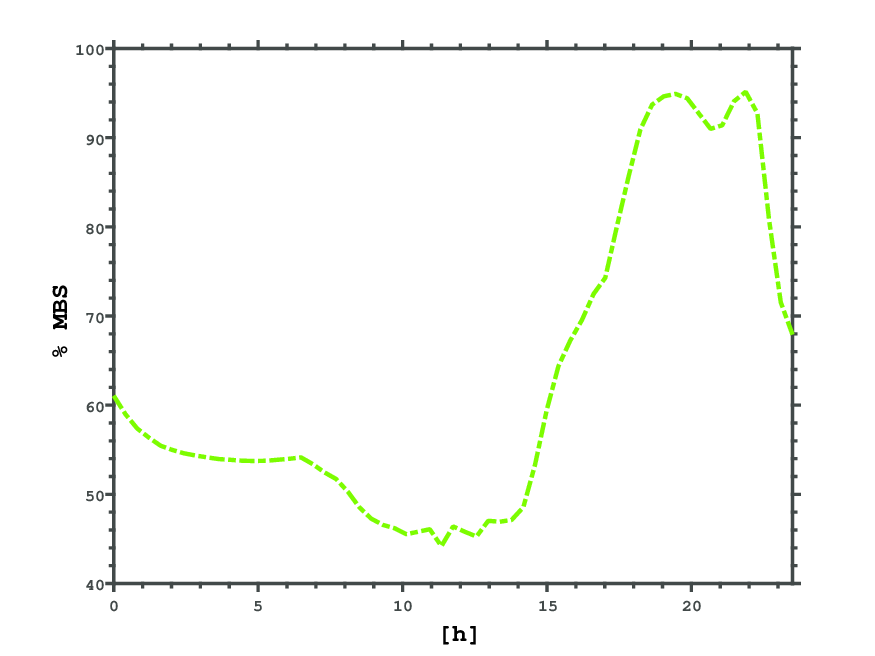}}
       \subfloat[]{\includegraphics[width=.3\linewidth]{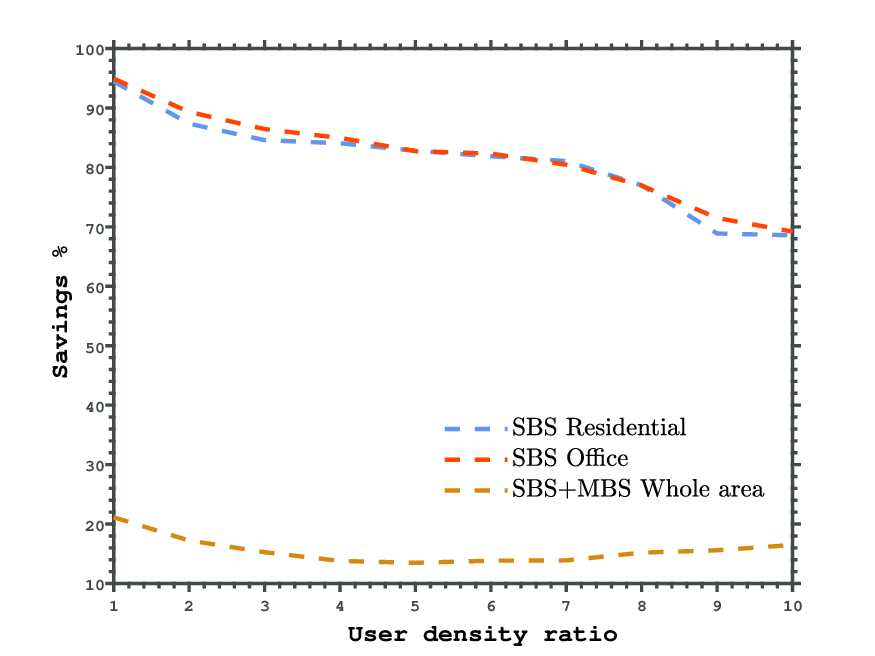}}
     \caption{(a) Fraction of the total amount of MBS in the scenario allocated to the residential area, over the $24$ hours; (b) Saving with respect to the fully static scenario, in terms of installed SBS in the residential and office districts and in terms of both moving and static BS in the whole area, as a function of the ratio between the values of maximum density of connected users in the two regions.}  
     \label{analysis}
     \vspace{-15pt}
\end{figure*}
We have derived the optimal number of MBSs in every time slot and the optimal total number of BSs in both regions, as derived from the solution of Problem 2, as well as the minimum number of BSs required to serve users in each region with the given target QoS, over 24 hours, as derived from the solution of Problem 1 (denoted as the "baseline"). As Fig. \ref{comparisons} shows,  
 the optimal density of \textit{active} BSs (baseline curve in Fig. 1b and 1c) is roughly directly proportional to the density of active users. The maximum of this quantity in each region (straight black line in Fig. 1b and 1c) is the number of BSs installed in the purely static scenario. Being dimensioned for peak traffic, the static scenario is over-provisioned during the majority of the day, with important implications in terms of consumed power. The plots show also how the adaptive densification brought in by MBSs substantially reduces the amount of redundant infrastructure during the 24h, potentially reducing also energy consumption. The plots show also that the maximum aggregate traffic demand in the two regions over the day ultimately determines the overall amount of static and moving BS deployed in the whole scenario. As the overall number of MBSs is driven by peak aggregate traffic demand, some amount of overprovisioning is present also in the population of MBSs, as visible in Fig. 1. 
 In realistic settings, this extra service capacity is however useful to account e.g., for a weak correlation between patterns of vehicle densification and patterns of network traffic demand. 
 On the other hand, our evaluation of the number of base stations in excess allows estimating the potential for energy savings achievable by adapting, in each time slot, the number of active (i.e. not in standby mode, and ready to serve users) base stations in each area, to have it coincide with the minimum required to serve users with the target QoS.\\  
Fig. 2a shows how the population of MBS is shared between the two regions at the optimum, as a function of the time of the day. The figure suggests that MBS are shared in an approximately equal manner between the two regions for most of the day. However, in the period during which network traffic demand peaks in the residential area, almost all of the MBSs converge there to supply the service capacity required. This strongly correlates with typical traffic patterns in an urban scenario, and further supports the feasibility of the moving base station paradigm in these settings.\\
One of the key system parameters that determine the optimal deployment strategies resulting from our approach is the density of connected users over the observation window. To make our results more robust, in a new set of experiments, we evaluated how the savings in the total amount of BS deployed enabled by the moving base station paradigm are impacted by this parameter. Specifically, we conduct our analysis by varying the ratio between the maximum density of connected users in the office district and the same quantity for the residential district (henceforth denoted as \textit{user density ratio}).
Adopting the traffic profiles in Fig. 1a, we set the peak at $10000$ users per $km^2$. Then we tuned the user density ratio letting the maximum density of connected users in the residential district vary. Fig. \ref{analysis}b shows the results for a user density ratio between 1 and 10. The figure indicates that in all configurations and areas, there is an average decrease of 17\% in the total number of base stations deployed, compared to the fully static configuration. This decrease can go up to 21\% for low values of user density ratio. 
Thus, the MBS paradigm, coupled with optimal MBS deployment strategies, can achieve the target user-perceived QoS while substantially decreasing the amount of static BS deployed, over a large spectrum of configurations in terms of user density ratio.
In addition, the reduction in terms of SBS deployed with respect to the static configuration varies between $95\%$ for low user density ratios, to $70\%$ for higher values. As expected, such a decrease is inferior with a larger user density ratio. Indeed, larger values of this parameter imply that residential active users are spread over the area, with a larger mean distance from the serving base station and a consequent decrease in the efficiency of the information transfer. Thus the base stations required to serve users increase, even in periods of low network traffic demand. As a consequence, we witnessed a reduction in the margins for MBS reuse in the two regions, and thus in the gains of the moving base stations paradigm. Anyway, the overall results show that MBS can remarkably decrease the dependence of the network on static deployments, which typically imply higher costs and must account for way tighter administrative and legal constraints than for MBS.\\
\vspace{-0.2in}
\section{Conclusions}
We presented a novel approach to evaluate the feasibility of the MBS paradigm and the potential resource savings it enables. The analysis suggests that MBSs effectively help reduce base station densification in high-density urban scenarios. Our results, though initial, are robust when varying the user density. Future works will include improved modeling of MBSs together with backhauling, and energy efficiency considerations.  
\vspace{-0.2in}

\section*{Acknowledgements}

This work was supported in part by the German Research Foundation (DFG) within the DyMoNet project under grant DR 639/25-1.
\vspace{-0.25in}

%


\end{document}